\RequirePackage[2020-02-02]{latexrelease}
\documentclass[superscriptaddress,showpacs,amsmath,amssymb,twocolumn,nofootinbib]{revtex4-1}
\usepackage{amsmath, amsthm, amssymb, bm, braket}
\usepackage[dvips]{graphicx}
\usepackage{color}
\newcommand{\kimu}[1]{\textcolor{blue}{#1}}

\begin{document}

%%%\title{Interplay of two $0^+$ resonances in two-proton emission of $^{16}$Ne}
\title{Interference of resonances in two-proton emission of $^{16}$Ne}

\author{Tomohiro Oishi}
\email[E-mail: ]{tomohiro.oishi@ribf.riken.jp}
\affiliation{RIKEN Nishina Center for Accelerator-Based Science, Wako 351-0198, Japan}

\author{Masaaki Kimura}
\email[E-mail: ]{masaaki.kimura@ribf.riken.jp}
\affiliation{RIKEN Nishina Center for Accelerator-Based Science, Wako 351-0198, Japan}

\renewcommand{\figurename}{FIG.}
\renewcommand{\tablename}{TABLE}

\newcommand{\bi}[1]{\ensuremath{\boldsymbol{#1}}}
\newcommand{\unit}[1]{\ensuremath{\mathrm{#1}}}
\newcommand{\oprt}[1]{\ensuremath{\hat{\mathcal{#1}}}}
\newcommand{\abs}[1]{\ensuremath{\left| #1 \right|}}
\newcommand{\slashed}[1] {\not\!{#1}} %--- e.g. \slashed{p} = \gamma^{\mu} p_{\mu}.

\newcommand{\crc}[1] {c^{\dagger}_{#1}}
\newcommand{\anc}[1] {c_{#1}}
\newcommand{\crb}[1] {\alpha^{\dagger}_{#1}}
\newcommand{\anb}[1] {\alpha_{#1}}

\def \beq{\begin{equation}}
\def \eeq{\end{equation}}
\def \beqa{\begin{eqnarray}}
\def \eeqa{\end{eqnarray}}

\def \bir{\bi{r}}
\def \ubir{\bar{\bi{r}}}
\def \bip{\bi{p}}
\def \ubip{\bar{\bi{r}}}
\def \adel{\tilde{l}} %--- orbital angular-momentum number for the small component, g(r).
\def \twop{$2p$}

\begin{abstract}  %\begin{description} \item[Background]
{\noindent
We investigate the two-proton ($2p$) emission from $^{16}$Ne with the time-dependent $^{14}$O$+p+p$ three-body calculations.
Two $0^+$ resonances are suggested to participate.
For the $0^+_1$ resonance, the true emission of the spatially localized two protons is dominant, if the single-particle $s_{1/2}$ resonance locates above the $2p$ energy.
By evaluating a time-evolution probability, the $0^+_2$ resonance is suggested at $\cong 3$ MeV from the $2p$ threshold.
\kimu{The inclusion of $0^+_2$ resonance makes the true-$2p$~emission more dominant in the time-dependent calculations.}
}
\end{abstract}

%%%\pacs{05.30.Fk, 21.10.Pc, 21.60.-n, 23.20.-g}
\maketitle

\section{Introduction} \label{sec:intro}
Two-proton (\twop) radioactivity is a characteristic
decay process beyond the proton-drip line \cite{2012Pfu_rev, 08Blank_01, 08Blank_02, 2009Gri_rev, 2023Pfutzner_rev, 2019Qi_rev}.
This phenomenon serves as a natural laboratory to investigate
the open-quantum systems, quantum entanglement,
and pairing correlations in nuclear systems.
From the energy release, lifetime, and \twop~correlations, we earn a deeper knowledge on these issues.
Such utility is unique in the proton-rich side \cite{2019Qi_rev}, whereas
the neutron-rich nuclei without Coulomb barriers rarely have the meta-stable states.
Because of the remarkable developments in the experimental
techniques \cite{2012Pfu_rev, 08Blank_01, 08Blank_02, 2009Gri_rev, 2023Pfutzner_rev},
information on \twop~emissions has been accumulated.

In the first prediction of \twop~emissions by Goldansky, two mechanisms were proposed;
simultaneous and sequential emissions \cite{60Gold, 61Gold}.
However in recent measurements, the actual process
manifests more complicated characters \cite{1989Boch, 2009Gri_rev, 12Ego}.
A combination of energy release and angular correlation of relative momenta
has been utilized to identify them \cite{08Muk, 10Muk, 12Ego, 01Gri_I, 10Gri_V, 09Gri_80, 09Gri_677, 2007Miernik}.
For such a complicated process, one cannot approximate it as the direct product of partial two-body problems, and thus, needs to deal with three particles without any discrimination.

The $^{16}$Ne nucleus is a \twop~emitter, which can be approximated as the two valence protons around the proton-shell-closed core $^{14}$O ($Z=8$).
Its subsystem $^{15}$F is unbound and has two low-lying resonances in the $s_{1/2}$ and $d_{5/2}$ channels \cite{91Ajzen, 2003Peters, 2004Goldberg, 2005Guo}.
Therefore, the proton-proton interaction
can induce the mixing of the $(s_{1/2})^2$ and $(d_{5/2})^2$ configurations, which will affect the \twop-emitting process of $^{16}$Ne.
In particular, the $s_{1/2}$ resonance in $^{15}$F can play a role of so-called intermediate resonance to invoke the sequential emission \cite{78KeKe, 2004Goldberg, 2005Guo, 2015Gri_16Ne}.
However, the recent experiments have shown that the decay of $^{16}$Ne is dominated by
the prompt (direct or true) \twop~emission from the lowest $0^+$ resonance \cite{2014Brown, 2016Charity_EPJCON}.
For its decay width, the early experiment gave $\Gamma_{2p}=0.11(4)$ MeV \cite{83Wood}.
Recent experiments have reported the possibility of narrower width \cite{2014Brown, 2014Wamers},
which is not in contradiction to theoretical predictions of keV-order width \cite{2002Grigorenko, 2014Fortune, 2015Gri_16Ne}.
On the other side, there can exist the second $0^+$ resonance \cite{1997Fohl}.
Whether this $0^+_2$ plays a sizable role or not in the emission process is still an open question.
Note that, in the isobaric-analogue partner $^{16}$C, the bound $0^+_2$ state has been expected \cite{1977Fortune}.
The co-existence of $0^+_1$ and $0^+_2$ states is common in other $sd$-shell even-even nuclei, e.g. $^{18}$Ne and $^{18}$O \cite{NNDC_Chart}.

This paper aims to clarify (i) whether the $0^+_2$ resonance exists or not in $^{16}$Ne, and (ii) how it affects the \twop-emitting process.
We employ the time-dependent three-body ($^{14}$O$+p+p$) calculation, where
both the $0^+_1$ and $0^+_2$ resonances can be involved.
The time-dependent calculation is probably the best option to visualize the behaviour of meta-stable state \cite{87Gur, 04Gur, 99Talou_60, 00Talou, 2014Oishi, 2017Oishi, 2021Wang_Naza, 2023SMWang},
and thus,
enables us to intuitively understood the \twop~dynamics by following their time evolution.
The decaying width or equivalently lifetime is directly evaluated from the time-dependent tunneling process.

In the next Sec. \ref{sec:form}, we give the formalism for time-dependent three-body calculations.
Sec. \ref{sec:resul} is devoted for presenting our results and discussions.
Finally we summarize this paper in Sec. \ref{sec:summary}.

\section{model and formalism} \label{sec:form}
We employ the three-body model, which has been developed and utilized in Refs. \cite{2005HS,07Hagi_01,07Bertulani_76,14Hagi_2n,88Suzuki_COSM,1991BE,1997EBH,2014Lorenzo,2010Oishi,2014Oishi,2017Oishi}.
The system contains an inert core ($^{14}$O) of mass $m_C$ and two valence protons with the following assumptions:
(i) the inert core is spherical and has $0^+$ spin and parity;
(ii) each valence proton interacts with the core through the spherical mean field $V$;
(iii) two protons interact through the potential $v_{pp}$.
Starting from the general coordinates $\left\{ \bi{x}_1, \bi{x}_2, \bi{x}_C \right\}$ of two protons and the core,
the total Hamiltonian reads
\beqa
\hat{H}_{3B} &=& -\frac{\hbar^2}{2m_p} \frac{\partial^2}{\partial \bi{x}^2_1} -\frac{\hbar^2}{2m_p} \frac{\partial^2}{\partial \bi{x}^2_2} -\frac{\hbar^2}{2m_C} \frac{\partial^2}{\partial \bi{x}^2_C} \nonumber \\
&&+V(\bi{x}_C,\bi{x}_1)+V(\bi{x}_C,\bi{x}_2)+v_{pp}(\bi{x}_1,\bi{x}_2),
\eeqa
where $m_p=938.272$ MeV$/c^2$ is the proton mass.
From these general coordinates, the core-orbital transformation is defined as
\beq
\left[ \begin{array}{l} \bir_1 \\ \bir_2 \\ \bi{R}_{CM} \end{array} \right]
=
\left( \begin{array}{ccc} 1 &0 &-1 \\ 0 &1 &-1 \\ \frac{m_p}{M} &\frac{m_p}{M}  &\frac{m_C}{M}  \end{array} \right)
\left[ \begin{array}{l} \bi{x}_1 \\ \bi{x}_2 \\ \bi{x}_C \end{array} \right],
\eeq
where $M=m_p+m_p+m_C$, and thus, $\bi{R}_{CM}$ is the center-of-mass coordinate.
With these new coordinates, the three-body Hamiltonian reads
\beq
\hat{H}_{3B} = \hat{h}(\bir_1)+\hat{h}(\bir_2) +v_{pp}(\bir_1,\bir_2) +\frac{\bip_1 \cdot \bip_2}{m_C} +\frac{\bi{P}^2_{CM}}{2M}. \label{eq:H3B}
\eeq
The center-of-mass term $\bi{P}^2_{CM}/2M$ is separated and omitted in the following.

%%%%%%%%%%%%%%%%%%%%%%%%%%%%%%%%%%%%%%%%%%%%%%%%%%%%%%%%%%%%%%%%%%%%%%%%%%%
\begin{table}[b] \begin{center}
\caption{Summary of model parameters.
}
\label{table:KIM}
\catcode`? = \active \def?{\phantom{0}} %define `?' as ' '(one-blank).
  \begingroup \renewcommand{\arraystretch}{1.2}
  \begin{tabular*}{\hsize} { @{\extracolsep{\fill}} llrr}
\hline \hline%%%%%%%%\multicolumn{3}{c}{(even-even)}
 &                                       &prompt  &mixed  \\  \hline
core-proton  &$r_0$~[fm]                 &$1.23$  &$1.23$   \\
             &$a_0$~[fm]                 &$0.64$   &$0.82$   \\
             &$V_0$~[MeV]                &$-42.25$ &$-51.53$    \\
             &$U_{ls}$~[MeV$\cdot$fm$^2$] &$54.1$   &$18.68$ \\
proton-proton  &$w_0$~[MeV$\cdot$fm$^3$]  &$-527$   &$-260$  \\
\hline \hline
  \end{tabular*}
  \endgroup
  \catcode`? = 12 %initialize `?'.
\end{center} \end{table}
%%%%%%%%%%%%%%%%%%%%%%%%%%%%%%%%%%%%%%%%%%%%%%%%%%%%%%%%%%%%%%%%%%%%%%%%%%%

%%%%%%%%%%%%%%%%%%%%%%%%%%%%%%%%%%%%%%%%%%%%%%%%%%%%%%%%%%%%%%%%%%%%%%%%%%%
\begin{table}[b] \begin{center}
\caption{Core-proton $s_{1/2}$ and $d_{5/2}$ resonances of the $^{15}$F = $^{14}$O$+p$.
The unit is MeV.
}
\label{table:GIS}
\catcode`? = \active \def?{\phantom{0}} %define `?' as ' '(one-blank).
  \begingroup \renewcommand{\arraystretch}{1.2}
  \begin{tabular*}{\hsize} { @{\extracolsep{\fill}} rllll}
\hline \hline
$^{15}$F~~  &\multicolumn{2}{c}{$s_{1/2}$} &\multicolumn{2}{c}{$d_{5/2}$}  \\
               &$E_{p} (s_{1/2})$      &$\Gamma_{p} (s_{1/2})$  &$E_{p} (d_{5/2})$      &$\Gamma_{p} (d_{5/2})$  \\  \hline
Expt.\cite{2005Guo}~     &$1.23(5)$     &$0.50$-$0.84$ &$2.81(2)$  &$0.30(6)$    \\
    \cite{2004Goldberg}~ &$1.23$-$1.37$ &$0.7$         &$2.795(45)$ &$0.325(6)$ \\
                        &$1.35$-$1.61$  &$-$  \\
    \cite{2003Peters}~   &$1.51(15)$   &$1.2$  &$2.853(45)$  &$0.34$  \\
    \cite{91Ajzen}~      &$1.47(13)$   &$1.0(2)$  &$2.77(10)$  &$0.24(3)$  \\
                prompt~  &$\cong 1.5$  &$\cong 1.3$  &$2.785 $ &$0.391$  \\
                 mixed~  &$1.277$      &$0.604$      &$2.787$  &$0.264$  \\
%%%\hline
%%%$^{15}$C,~Expt. \cite{NNDC_Chart}  &$-1.2181(8)$ &$-$  &$-0.4780(15)$  &$-$  \\
%%% This work, prompt  &$-0.238$  &$-$  &$-0.147$ &$-$  \\
%%%            mixed  &$-1.257$  &$-$  &$-0.501$ &$-$  \\
\hline \hline
  \end{tabular*}
  \endgroup
  \catcode`? = 12 %initialize `?'.
\end{center} \end{table}
%%%%%%%%%%%%%%%%%%%%%%%%%%%%%%%%%%%%%%%%%%%%%%%%%%%%%%%%%%%%%%%%%%%%%%%%%%%

\subsection{core-proton subsystem}
In Eq. (\ref{eq:H3B}), $\hat{h}(\bir_i)$ is the Hamiltonian for the subsystem composed of the $i$th valence proton and the core:
\beqa
&& \hat{h}(r_i) = -\frac{\hbar^2}{2\mu} \frac{d^2}{dr_i^2} + V(r_i), \nonumber \\
&& V(r_i) = \frac{\hbar^2}{2\mu} \frac{l(l+1)}{r_i^2} +V_{WS}(r_i) +V_{Coul}(r_i),
\eeqa
where $\mu = m_p m_C /(m_p + m_C)$.
To describe this subsystem, $^{15}$F$= ^{14}$O$+p$,
we employ the spherical Woods-Saxon and Coulomb potentials.
The Woods-Saxon potential reads
\beqa
&& V_{WS}(r) = V_0 f(r) + U_{ls} (\bi{l} \cdot \bi{s}) \frac{1}{r} \frac{df(r)}{dr}, \nonumber  \\
&& f(r) = \frac{1}{1 + e^{(r-R_0)/a_0}}, \label{eq:FWS15}
\eeqa
where $R_0=r_0\cdot 14^{1/3}$ and $r_0 = 1.23$ fm.
Other parameters given in TABLE \ref{table:KIM} will be discussed later.
In addition, the Coulomb potential of uniformly-charged sphere with radius $R_0$ is employed:
\beq
V_{Coul}(r)= \left\{ \begin{array}{ll}
  \displaystyle \hbar c \frac{Z \alpha_0}{r} &(r \ge R_0) \\
  \displaystyle \hbar c \frac{Z \alpha_0}{2R_0} \left[ 3-\left(\frac{r}{R_0} \right)^2 \right]  &(r<R_0) \end{array} \right. ,
\eeq
where $\alpha_0 = \frac{e^2}{4\pi \epsilon_0 \hbar c} \cong \frac{1}{137.036}$ and $Z=8$.
The Sch\"{o}dinger equation for the core-proton subsystem reads
\beq
\hat{h} \ket{\phi_a} = e_a \ket{\phi_a},
\eeq
where $e_a$ is the eigenenergy of the single-particle level with
the quantum numbers $a=\{ n_a, l_a, j_a, m_a \}$,
which denote the nodal quantum number, orbital angular momentum, coupled angular momentum, and
magnetic quantum number, respectively.
The single-particle wave function thus reads
\beq
\phi_a(\bir \sigma) =R_{n_a l_a j_a}(r) \left[ Y_{l_a}(\theta, \phi) \otimes \chi_{\frac{1}{2}} (\sigma) \right]^{(j_a m_a)}.
\eeq
The radial part $R_{n_a l_a j_a}(r)$ is numerically solved with Numerov method.
The potential $V(r)$ has the bound solutions of $0s_{1/2}$, $0p_{3/2}$, and $0p_{1/2}$ states.
These bound states are excluded from the model space for the three-body Hamiltonian, due to the Pauli principle.
The continuum states with $e_a >0$ are discretized within a box of $r_{\rm max}=80$ fm radius.

%%%%%%%%%%%%%%%%%%%%%%%%%%%%%%%%%%%%%%%%%%%%%%%%%%%%%%%%%%%%%%%%%%%%%%
\begin{figure}[t] \begin{center}
\includegraphics[width = \hsize]{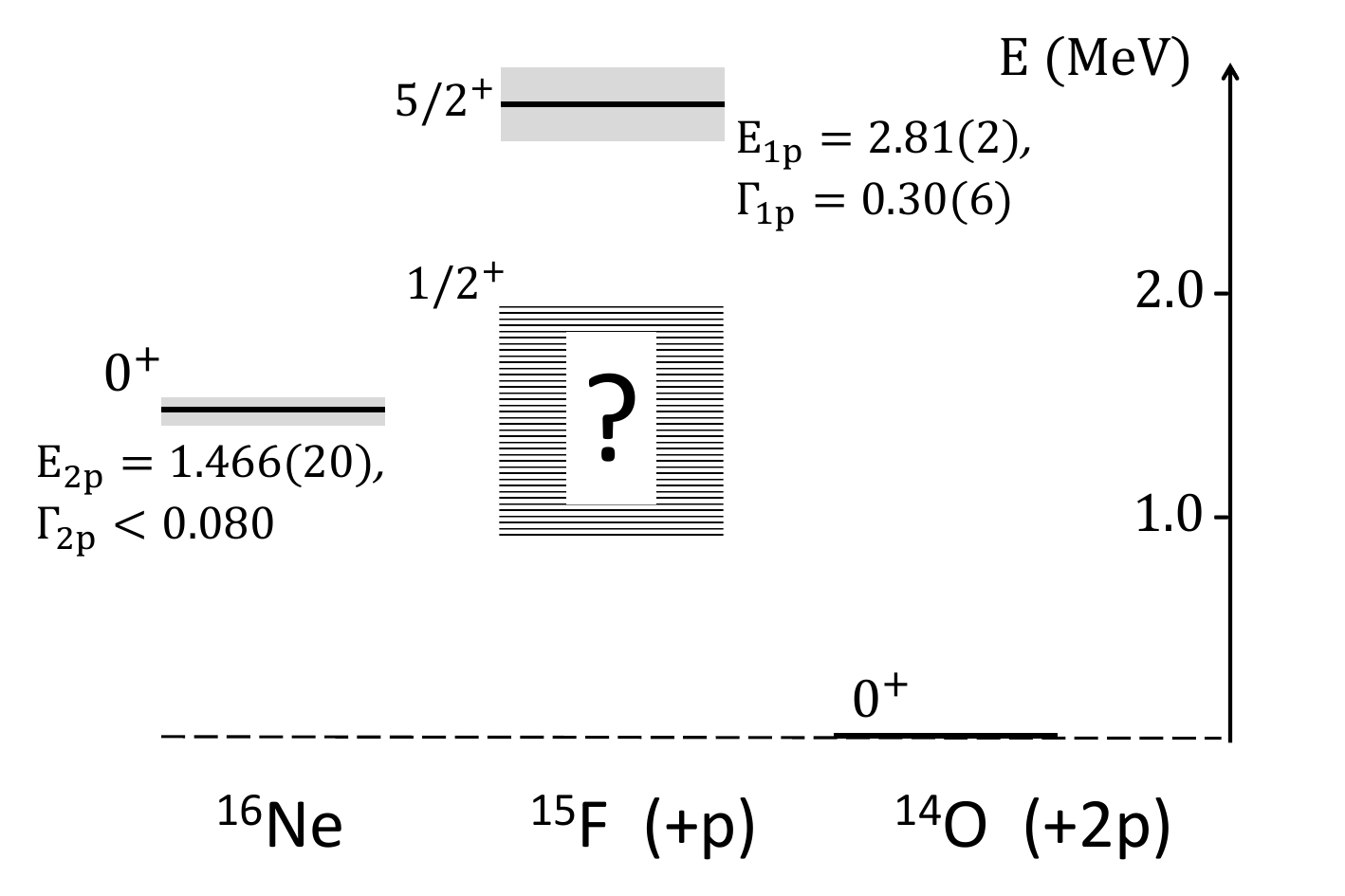}
\caption{Level scheme of the $^{16}$Ne, $^{15}$F, and $^{14}$O nuclei.
Data are taken from Ref. \cite{2005Guo} for the $d_{5/2}$ of $^{15}$F and
Ref. \cite{2014Brown} for $^{16}$Ne.
For the $s_{1/2}$ of $^{15}$F, several sets of experimental data exist, and those are summarized in TABLE \ref{table:GIS}.
}\label{fig:2023_0724}
\end{center} \end{figure}
%%%%%%%%%%%%%%%%%%%%%%%%%%%%%%%%%%%%%%%%%%%%%%%%%%%%%%%%%%%%%%%%%%%%%%

%%%%%%%%%%%%%%%%%%%%%%%%%%%%%%%%%%%%%%%%%%%%%%%%%%%%%%%%%%%%%%%%%%%%%%
\begin{figure}[t] \begin{center}
\includegraphics[width = \hsize]{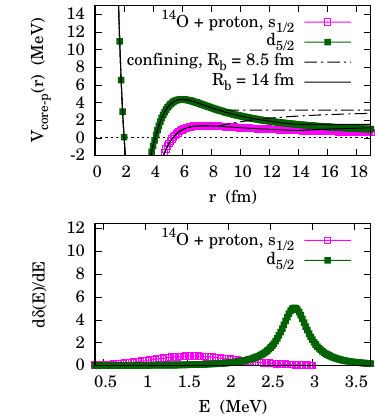}
\caption{(Top) Core-proton potentials for the $^{14}$O+$p$ subsystem in the prompt case.
The confining potentials for time-dependent calculations are also plotted.
(Bottom) Derivative of scattering phase shift obtained with the core-proton potentials.
}
\label{fig:2023_1107}
\end{center} \end{figure}
%%%%%%%%%%%%%%%%%%%%%%%%%%%%%%%%%%%%%%%%%%%%%%%%%%%%%%%%%%%%%%%%%%%%%%
%%%%%%%%%%%%%%%%%%%%%%%%%%%%%%%%%%%%%%%%%%%%%%%%%%%%%%%%%%%%%%%%%%%%%%
\begin{figure}[t] \begin{center}
\includegraphics[width = \hsize]{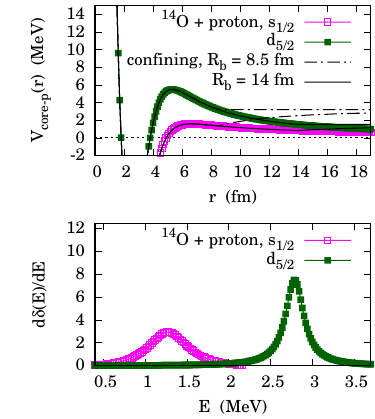}
\caption{Same to FIG. \ref{fig:2023_1107} but in the mixed case.
}
\label{fig:2023_0627}
\end{center} \end{figure}
%%%%%%%%%%%%%%%%%%%%%%%%%%%%%%%%%%%%%%%%%%%%%%%%%%%%%%%%%%%%%%%%%%%%%%

In TABLE \ref{table:KIM},
parameters for $V(r_i)$ are summarized.
We remark that there is an ambiguity in the $s_{1/2}$ resonance \cite{2003Peters, 2005Guo, 2015Gri_16Ne}.
Several experiments have suggested different energies and widths, which are summarized in TABLE \ref{table:GIS}.
In Ref. \cite{2003Peters}, the $s_{1/2}$ resonance is reported at $1.51(15)$ MeV, being higher than the $0^+_1$ resonance of $^{16}$Ne.
The first set of parameters denoted by ``prompt'' reproduces these data \cite{2003Peters}.
The calculated results are summarized in TABLE \ref{table:GIS}.
In FIG. \ref{fig:2023_1107}, the actual shape of core-proton potential $V(r)$ and
the energy derivative of the scattering phase shift, $d \delta /dE$, are presented.
Notice that the potential barrier exists around $r \cong 5$ fm.
It becomes higher in the $d_{5/2}$ channel, because of the centrifugal term, $\frac{\hbar^2}{2\mu} \frac{l(l+1)}{r^2}$.

On the other hand,
in Ref. \cite{2005Guo}, the $s_{1/2}$ resonance is reported at $1.23(5)$ MeV.
This is lower than the $0^+_1$ resonance of $^{16}$Ne, and thus, can play a role of the intermediate resonance.
We introduce the second setting denoted by ``mixed'' which reproduces these data as shown in FIG. \ref{fig:2023_0627} and TABLE \ref{table:GIS}.
As explained in the next section,
the energy of $s_{1/2}$ resonance remarkably affects the results of the time-development calculations.

\subsection{two-proton state}
We define the no-correlation \twop~states as
\beq
\tilde{\Phi}^{(JM,\pi)}_{ab}(\bir_1 \sigma_1 ,\bir_2 \sigma_2) = \hat{A} \left[ \phi_a (\bir_1 \sigma_1) \otimes \phi_b (\bir_2 \sigma_2) \right]^{(JM,\pi)},
\eeq
where $J=\abs{\bi{j}_a +\bi{j}_b}$ (coupled angular momentum), $M=0,\pm 1,\cdots,\pm J$ (magnetic quantum number), and $\pi=(-)^{l_a +l_b}$ (parity).
Here $\hat{A}$ indicates the anti-symmetrization.
For the three-body calculations, we employ the single-particle orbits up to $l_{\rm cut}=7$ with the cutoff energy, $E_{\rm cut}=24$ MeV.
In the following, we discuss the $J=M=0$ and $\pi=+$, and hence, omit these labels.
Notice that $\tilde{\Phi}_{ab}(\bir_1 \sigma_1 ,\bir_2 \sigma_2)$ are the eigenstates of the no-correlation Hamiltonian without the proton-proton interaction and recoil term:
\beq
\left[ \hat{h}(\bir_1) +\hat{h}(\bir_2) \right] \tilde{\Phi}_{ab} = \left( e_a +e_b\right) \tilde{\Phi}_{ab}.
\eeq
By using these states as bases,
the three-body eigenstate $\ket{E_N}$, which satisfies $\hat{H}_{3B} \ket{E_N}= E_N \ket{E_N}$, can be expanded as
\beq
\Braket{\bir_1 \sigma_1 ,\bir_2 \sigma_2 | E_N} = \sum_{ab} U_{N,ab} \tilde{\Phi}_{ab}(\bir_1 \sigma_1 ,\bir_2 \sigma_2),
\eeq
where the expanding coefficients $U_{N,ab}$ are determined by diagonalizing the
Hamiltonian given by Eq. (\ref{eq:H3B}).

The proton-proton interaction includes the vacuum and additional, surface-dependent terms:
\beq
v_{pp}(\bir_1,\bir_2) = v_{pp,vac}(\bir_1,\bir_2) + v_{pp,add}(\bir_1,\bir_2).
\eeq
The vacuum potential includes the nuclear and Coulomb terms \cite{2014Oishi}.
Namely,
\beqa
&& v_{pp,vac}(\bir_1,\bir_2) = \Bigl[ V_R e^{-a_R r^2} +V_S e^{-a_S r^2} \hat{P}_{S=0} \Bigr. \nonumber  \\
&& ~~~~~~~\Bigl. +V_T e^{-a_T r^2} \hat{P}_{S=1} \Bigr] +v_{pp,Coul}(r),  \label{eq:vpptue}
\eeqa
where $r=\abs{\bir_2 -\bir_1}$ and $v_{pp,Coul}(r)=\hbar c  \alpha_0 /r$.
The operators $\hat{P}_{S=0}$ and $\hat{P}_{S=1}$ are the projectors to the spin-singlet and spin-triplet channels, respectively.
Here the parameters are set as
$V_R=200$ MeV,
$V_S=-91.85$ MeV,
$V_T=-178$ MeV,
$a_R=1.487$ fm$^{-2}$,
$a_S=0.465$ fm$^{-2}$, and $a_T=0.639$ fm$^{-2}$
to reproduce the experimental spin-singlet $s$-wave scattering of two protons \cite{77Thom}.

%%%%%%%%%%%%%%%%%%%%%%%%%%%%%%%%%%%%%%%%%%%%%%%%%%%%%%%%%%%%%%%%%%%%%%
\begin{figure}[t] \begin{center}
\includegraphics[width = \hsize]{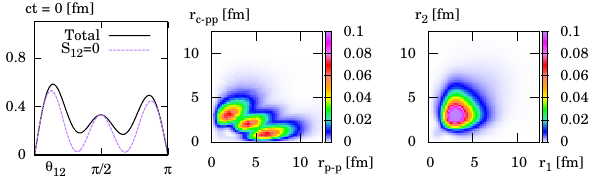}
  (a) $R_b = 14$~fm
\includegraphics[width = \hsize]{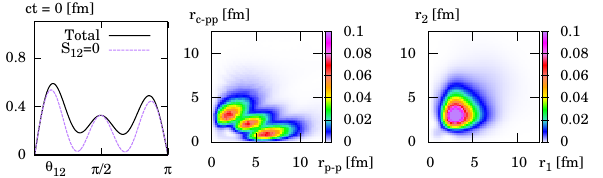}
  (b) $R_b = 8.5$~fm
\caption{\kimu{(a)
Density distribution of \twop~state at $t=0$ for $^{16}$Ne in the prompt case with $R_b =14$ fm.}
For convention of plotting variables, see Refs. \cite{2014Oishi, 2017Oishi, 2023Pfutzner_rev}.
(b) Same but by changing the confining radius as $R_b =8.5$ fm.}
\label{fig:GGG71_WS2}
\end{center} \end{figure}
%%%%%%%%%%%%%%%%%%%%%%%%%%%%%%%%%%%%%%%%%%%%%%%%%%%%%%%%%%%%%%%%%%%%%%

With only $v_{pp,vac}(\bir_1 ,\bir_2)$, the three-body model over-estimates
the observed \twop-energy, $Q_{2p,~{\rm expt.}}=1.401(20)$ MeV of $0^+_1$ \cite{2021Wang_AME}.
This indicates that the effective proton-proton interaction around the
core surface, i.e., in the dilute-density region, should be more attractive \cite{05Yama,2005Matsuo}.
Thus, we introduce the additional potential:
\beq
v_{pp,add} (\bir_1,\bir_2) = w_0 e^{-(R-R_0)^2 /B_0^2} \delta (\bir_1-\bir_2), \label{eq:vppadd}
\eeq
which is the contact-type interaction \cite{1991BE, 1997EBH},
whose strength depends on the radial profile to phenomenologically describe the attraction around the core surface $R_0$.
The parameters are set as $R=\abs{(\bir_1 +\bir_2) /2}$,
$R_0 = r_0 \cdot 14^{1/3}$, and
$B_0 = 0.6 R_0$.
The strength is adjusted to reproduce the observed $Q_{2p,~{\rm expt.}}$ as
$w_0 = -527$ ($-260$) MeV$\cdot$fm$^3$ for the prompt (mixed) case.
Notice that $v_{pp,add}$ becomes zero when one of the three particles is separated.
Therefore, this potential does not harm the two-body subsystems in vacuum.

%%%%%%%%%%%%%%%%%%%%%%%%%%%%%%%%%%%%%%%%%%%%%%%%%%%%%%%%%%%%%%%%%%%%%%
\begin{figure}[t] \begin{center}
\includegraphics[width = \hsize]{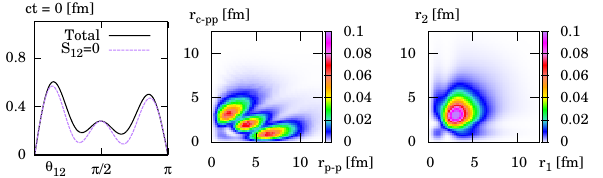}
 (a) $R_b = 14$~fm
\includegraphics[width = \hsize]{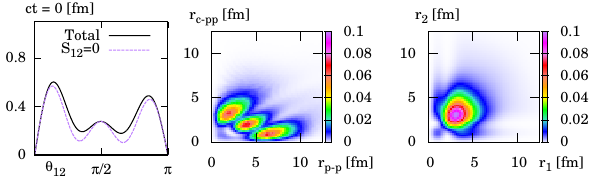}
 (b) $R_b = 8.5$~fm
\caption{Same to FIG. \ref{fig:GGG71_WS2} but in the mixed case.}
\label{fig:GGG71_WS1}
\end{center} \end{figure}
%%%%%%%%%%%%%%%%%%%%%%%%%%%%%%%%%%%%%%%%%%%%%%%%%%%%%%%%%%%%%%%%%%%%%%

%%%%%%%%%%%%%%%%%%%%%%%%%%%%%%%%%%%%%%%%%%%%%%%%%%%%%%%%%%%%%%%%%%%%%%%%%%%
\begin{table}[b] \begin{center}
\caption{
Properties of the initial \twop~state for the time-dependent calculations of $^{16}$Ne.
The \twop~energy $Q_{2p}$, expectation values of the proton-proton interactions, and the ratios of the $(s_{1/2})^2$ and $(d_{5/2})^2$ components are presented.
}
\label{table:POX}
\catcode`? = \active \def?{\phantom{0}} %define `?' as ' '(one-blank).
\begingroup \renewcommand{\arraystretch}{1.2}
\begin{tabular*}{\hsize} { @{\extracolsep{\fill}} l rr c rr}
\hline \hline
          &\multicolumn{2}{c}{prompt}  &~~  &\multicolumn{2}{c}{mixed}   \\
$R_b$~[fm]                           &$14.0$    &$8.5$  &  &$14.0$    &$8.5$   \\   \hline
$Q_{2p}=\Braket{\hat{H}_{\rm 3B}}$~[MeV] &$1.404$   &$1.407$ &  &$1.401$   &$1.407$  \\
$\Braket{v_{pp}}$~[MeV]                &$-6.745$  &$-6.338$ &  &$-4.663$  &$-4.384$ \\
$\Braket{v_{pp,Coul}}$~[MeV]            &$0.462$   &$0.464$  &  &$0.453$   &$0.463$  \\
$\Braket{v_{pp,add}}$~[MeV]            &$-3.044$   &$-2.921$&   &$-1.981$   &$-1.707$  \\
$(s_{1/2})^2$ ratio~[\%]               &$8.5$    &$7.8$   &   &$32.0$     &$28.6$  \\
$(d_{5/2})^2$ ratio~[\%]               &$88.4$   &$89.2$  &   &$63.5$     &$66.9$  \\
\hline \hline
\end{tabular*}
\endgroup
\catcode`? = 12 %initialize `?'.
\end{center} \end{table}
%%%%%%%%%%%%%%%%%%%%%%%%%%%%%%%%%%%%%%%%%%%%%%%%%%%%%%%%%%%%%%%%%%%%%%%%%%%

\section{Results} \label{sec:resul}
We calculate the time development of the $^{14}$O$+p+p$ system with Hamiltonian $\hat{H}_{3B}$.
\kimu{For simulating the emission,
two protons should be phenomenologically confined inside the potential barrier of the core.}
To prepare such an initial \twop~state, $\ket{\Psi(t=0)}$,
we employ the confining-potential method \cite{87Gur,88Gur,94Serot,94Car,98Talou,00Talou,04Gur}.
Namely, a wall potential is introduced
\kimu{at $r \geq R_{b}$
with the confining radius, $R_{b} =14$ fm,}
at $t\le 0$ as displayed in FIGs. \ref{fig:2023_1107} and \ref{fig:2023_0627}.
In TABLE \ref{table:POX}, properties of the initial states are summarized.

%%%%%%%%%%%%%%%%%%%%%%%%%%%%%%%%%%%%%%%%%%%%%%%%%%%%%%%%%%%%%%%%%%%%%%
\begin{figure}[t] \begin{center}
\includegraphics[width = \hsize]{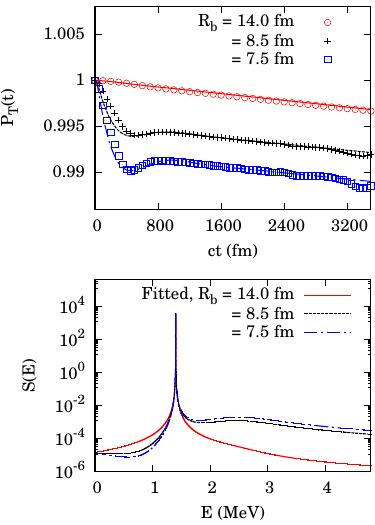}
\caption{\kimu{(Top)
Time-evolution probability $P_{T}(t)$ for $^{16}$Ne$\longrightarrow ^{14}$O$+p+p$ in the prompt case.
The $R_{b}$ indicates the confining radius.
(Bottom) Energy distribution corresponding to the fitted results of $P_{T}(t)$.
See Eq. (\ref{eq:TWOBW}) for definition.}}
\label{fig:2023_1113}
\end{center} \end{figure}
%%%%%%%%%%%%%%%%%%%%%%%%%%%%%%%%%%%%%%%%%%%%%%%%%%%%%%%%%%%%%%%%%%%%%%
%%%%%%%%%%%%%%%%%%%%%%%%%%%%%%%%%%%%%%%%%%%%%%%%%%%%%%%%%%%%%%%%%%%%%%
\begin{figure}[t] \begin{center}
\includegraphics[width = \hsize]{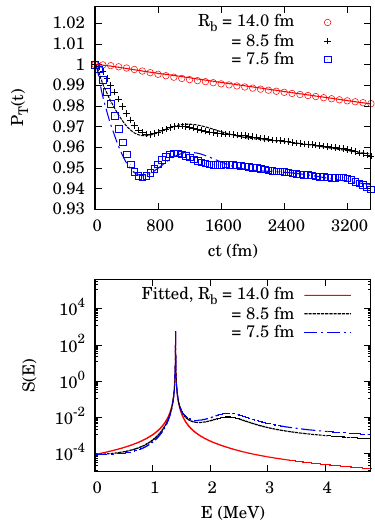}
\caption{\kimu{Same to FIG. \ref{fig:2023_1113} but in the mixed case.}}
\label{fig:2023_0618}
\end{center} \end{figure}
%%%%%%%%%%%%%%%%%%%%%%%%%%%%%%%%%%%%%%%%%%%%%%%%%%%%%%%%%%%%%%%%%%%%%%

Figures \ref{fig:GGG71_WS2} and \ref{fig:GGG71_WS1} display the density distributions of two protons at $t=0$.
In these figures and hereafter, we employ the following variables to describe the coordinates of the three-body system;
the \twop-opening angle $\theta_{12}$, and the distances between constituent particles,
$r_{\rm p-p} \equiv \abs{\bir_2 -\bir_1}$ and $r_{\rm c-pp} \equiv \abs{\bir_2 +\bir_1}/2$ as in Refs. \cite{2014Oishi, 2017Oishi, 2023Pfutzner_rev}.
One can see that the two protons are confined around the core.
\kimu{For both of the prompt and mixed cases,
the initial \twop~density distributions show a similar profile.}
Their angular distributions are asymmetric, where the peak at $\theta_{12} \leq \pi /4$ corresponds to the diproton-like localization.
As well known, this asymmetric form is a product of the mixture of $(l \in even)^2$ and $(l \in odd)^2$ configurations.
We note that the bound two-neutron ($2n$) state in $^{16}$C \cite{2005HS, 07Hagi_03}
shows a similar distribution to the present result.
%%%\tomo{Hence, in other words, we investigate the time evolution of $^{14}$O+p+p starting from the initial state similar to the ground state of $^{16}$C.}

%%%%%%%%%%%%%%%%%%%%%%%%%%%%%%%%%%%%%%%%%%%%%%%%%%%%%%%%%%%%%%%%%%%%%%%%%%%
\begin{table}[b] \begin{center}
\caption{Energy, width, and amplitudes of the $(d_{5/2})^2$ and $(s_{1/2})^2$ configurations of the $0^+_1$ and $0^+_2$ resonances participating in the time-dependent \twop~emission from $^{16}$Ne.
The $0^+_2$ width is energy-dependent: see Eq. (\ref{eq:Gofe2}) for definition.
Note that $u_2 =0$ when $R_b =14$ fm.
} \label{table:VEY}
\catcode`? = \active \def?{\phantom{0}} %define `?' as ' '(one-blank).
\begingroup \renewcommand{\arraystretch}{1.2}
\begin{tabular*}{\hsize} { @{\extracolsep{\fill}} llll }
\hline \hline
&~~~~~$k$               &$1$ for $0^+_1$    &$2$  for $0^+_2$ \\
\hline
prompt &$E_k$~[MeV]     &$1.404$             &$2.96$ \\
&$\Gamma_k$~[MeV]      &$1.4\times 10^{-4}$   &$2.77$ \\
&$u_2$($R_b=8.5$ fm)             &           &$-0.047$ \\
&$u_2$($R_b=7.5$ fm)             &           &$-0.061$ \\
&$(s_{1/2})^2$~[\%]        &$?8.5$            &$70.1$  \\
&$(d_{5/2})^2$~[\%]        &$88.4$            &$?5.7$  \\
\hline
mixed &$E_k$~[MeV]               &$1.401$       &$2.44$ \\
&$\Gamma_k$~[MeV]          &$1.1\times 10^{-3}$  &$0.82$ \\
&$u_2$($R_b=8.5$ fm)             &                 &$-0.109$ \\
&$u_2$($R_b=7.5$ fm)             &                 &$-0.140$ \\
&$(s_{1/2})^2$~[\%]    &$32.0$    &$61.2$  \\
&$(d_{5/2})^2$~[\%]    &$63.5$    &$33.7$  \\
\hline \hline
\end{tabular*}
\endgroup
\catcode`? = 12 %initialize `?'.
\end{center} \end{table}
%%%%%%%%%%%%%%%%%%%%%%%%%%%%%%%%%%%%%%%%%%%%%%%%%%%%%%%%%%%%%%%%%%%%%%%%%%%

\kimu{In FIGs \ref{fig:GGG71_WS2} and \ref{fig:GGG71_WS1}, the same results but by changing the confining radius, $R_b = 8.5$ fm, are also presented.
We mention these results in section \ref{sec:ddd}.
}

When the initial state is fixed, one can expand it as $\ket{\Psi(0)} = \sum_{N} F(0) \ket{E_N}$,
and its time development is solved as
\beq
\ket{\Psi(t)} = \exp \left[ -it \frac{\hat{H}_{3B}}{\hbar} \right] \ket{\Psi(0)} = \sum_{N} F_N(t) \ket{E_N},
\eeq
where $F_N(t)=e^{-it\frac{E_N}{\hbar}} F_N(0)$.
\kimu{Time-evolution probability is determined as $P_{T}(t) = \abs{\beta(t)}^2$, where \cite{47Krylov, 89Kuku}
\beq
  \beta(t) = \Braket{\Psi(0) | \Psi(t)} = \sum_{N} e^{-it\frac{E_N}{\hbar}} \abs{F_N (0)}^2.
\eeq
Note that $\beta(t) = \int e^{-it\frac{E}{\hbar}} \abs{F (E,t=0)}^2 dE$ in the continuous limit.
The mean-energy release, $Q_{2p}=-S_{2p}$, is evaluated as
\beqa
Q_{2p}= \Braket{\Psi(t) | \hat{H}_{\rm 3B} | \Psi(t)} = \sum_{N} \abs{F_N(0)}^2 E_{N}.
\eeqa
Notice that this quantity is time-independent.
}

%%%%%%%%%%%%%%%%%%%%%%%%%%%%%%%%%%%%%%%%%%%%%%%%%%%%%%%%%%%%%%%%%%%%%%
\begin{figure}[t] \begin{center}
\includegraphics[width = \hsize]{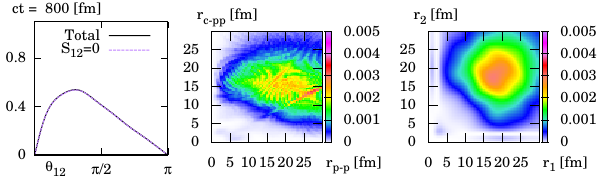}
\includegraphics[width = \hsize]{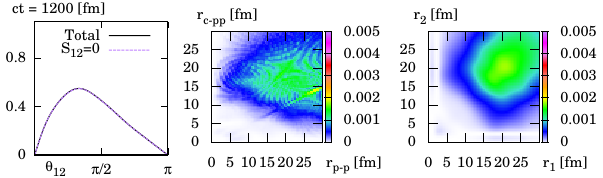}
\includegraphics[width = \hsize]{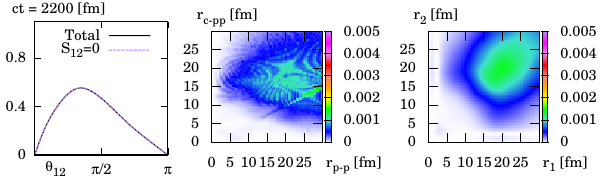}
\caption{
Density distributions of the time-dependent decaying \twop~state: see Eqs. (\ref{eq:CUSYGE}) and (\ref{eq:DAQUES}) for definition.
Results in the prompt case with $R_b =14$ fm.
}
\label{fig:COKG_WS2_GS}
\includegraphics[width = \hsize]{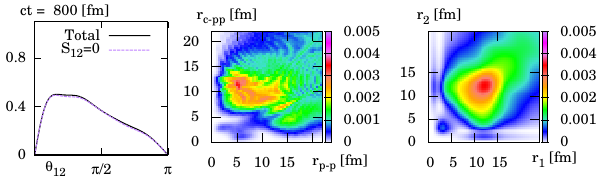}
\includegraphics[width = \hsize]{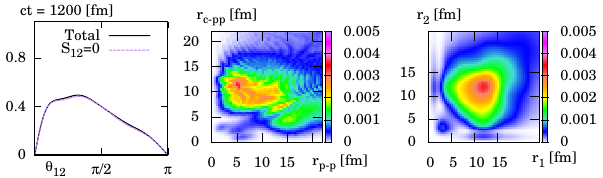}
\includegraphics[width = \hsize]{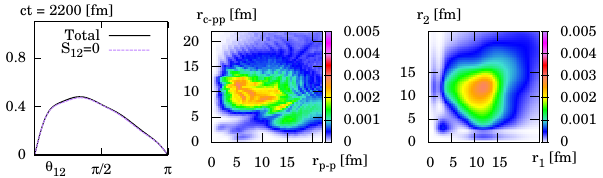}
\caption{
Same to FIG. \ref{fig:COKG_WS2_GS} (prompt) but $R_b =8.5$ fm.}
\label{fig:COKG_WS2}
\end{center} \end{figure}
%%%%%%%%%%%%%%%%%%%%%%%%%%%%%%%%%%%%%%%%%%%%%%%%%%%%%%%%%%%%%%%%%%%%%%

%%%%%%%%%%%%%%%%%%%%%%%%%%%%%%%%%%%%%%%%%%%%%%%%%%%%%%%%%%%%%%%%%%%%%%
\begin{figure}[t] \begin{center}
\includegraphics[width = \hsize]{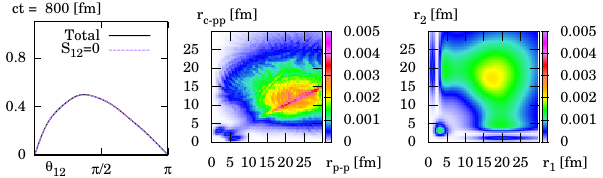}
\includegraphics[width = \hsize]{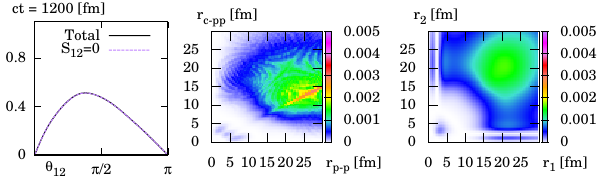}
\includegraphics[width = \hsize]{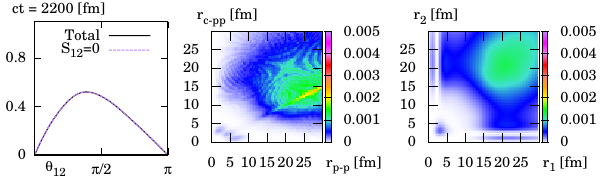}
\caption{
Same to FIG. \ref{fig:COKG_WS2_GS} but in the mixed case with $R_b =14$ fm.}
\label{fig:COKG_WS1_GS}
\includegraphics[width = \hsize]{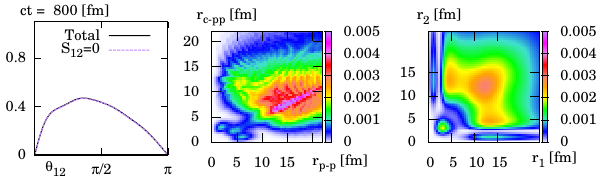}
\includegraphics[width = \hsize]{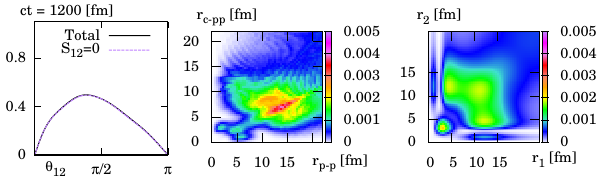}
\includegraphics[width = \hsize]{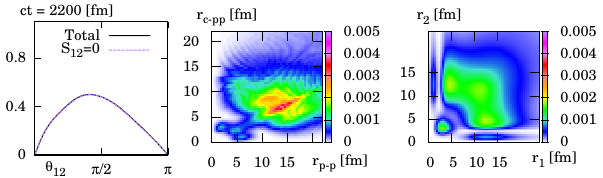}1
\caption{
Same to FIG. \ref{fig:COKG_WS1_GS} (mixed) but $R_b =8.5$ fm.}
\label{fig:COKG}
\end{center} \end{figure}
%%%%%%%%%%%%%%%%%%%%%%%%%%%%%%%%%%%%%%%%%%%%%%%%%%%%%%%%%%%%%%%%%%%%%%

%%%%%%%%%%%%%%%%%%%%%%%%%%%%%%%%%%%%%%%%%%%%%%%%%%%%%%%%%%%%%%%%%%%%%%
\begin{figure}[t] \begin{center}
    \includegraphics[width = 0.9\hsize]{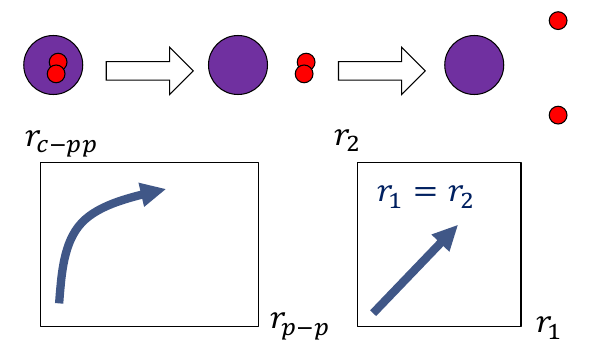}\\ true \twop~emission\\
    \vspace{12pt}
    \includegraphics[width = 0.9\hsize]{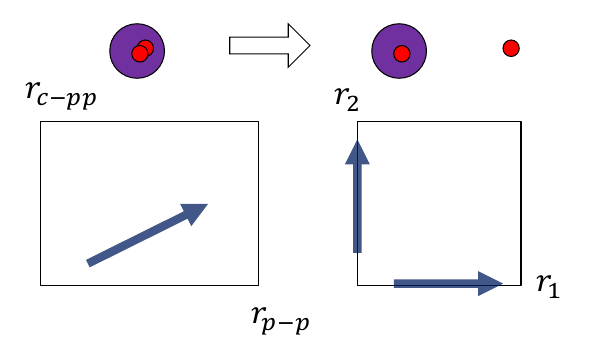}\\ sequential emission\\
    \vspace{12pt}
    \includegraphics[width = 0.9\hsize]{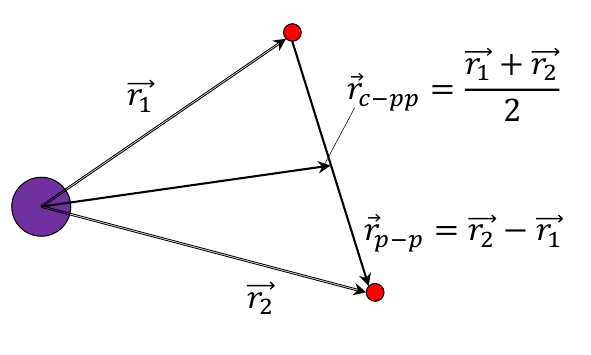}\\
\caption{Schematic pictures of the time-dependent
density distributions of the true \twop~emission (top)
and sequential $1p$-$1p$ emission (middle).
Definition of the coordinates are presented in the bottom panel.
}\label{fig:MAPS}
\end{center}
\end{figure}
%%%%%%%%%%%%%%%%%%%%%%%%%%%%%%%%%%%%%%%%%%%%%%%%%%%%%%%%%%%%%%%%%%%%%%

\subsection{first $0^+$ resonance}
First we evaluate the $0^+_1$-resonance energy and width.
For this purpose, the time-evolution probability from the initial state is calculated.
Note that $P_{T}(t) \cong e^{-t/\tau}$,
if the decay process was governed by a single-peak spectrum with Breit-Wigner profile, whose width is $\Gamma=\hbar / \tau$ \cite{89Kuku}.

The calculated $P_{T}(t)$ are shown in FIGs. \ref{fig:2023_1113} and \ref{fig:2023_0618}.
With $R_b = 14$ fm for the confining potential,
the $P_{T}(t)$ can be well approximated as the exponential damping both in the prompt and mixed cases.
Thus, the corresponding energy distribution has the single Breit-Wigner profile.
That is, for $\ket{\Psi(t=0)} =\int  F(E) \ket{E} dE$,
\beq
F(E)= L_1 (E) = \sqrt{\frac{\Gamma_1}{2 \pi}} \frac{1}{E-(E_1 -i \Gamma_1 /2)}. \label{eq:ONEBW}
\eeq
From numerical fitting, $E_1$ and $\Gamma_1$ are obtained as in TABLE \ref{table:VEY}.
The energy spectra, $S(E) = \abs{F(E)}^2$, are displayed in FIGs. \ref{fig:2023_1113} and \ref{fig:2023_0618}.
This narrow \twop~resonance at $1.4$ MeV coincides with the measured ground state of $^{16}$Ne.
Its \twop-emission width is in accordance with Ref. \cite{2014Brown}, which reports the upper limit of $\Gamma_{2p} \le 80$ keV, as well as with other theoretical predictions of keV-order width \cite{2002Grigorenko, 2014Fortune, 2015Gri_16Ne}.

Next we address the time-dependent motion of protons. %%%$\rho_d(t,\bir_1,\bir_2) =\abs{\Psi_d(t,\bir_1,\bir_2)}^2$.
For this purpose, the decaying state $\ket{\Psi_d(t)}$ is defined as the state orthogonal to the initial state:
\beq
\ket{\Psi_d(t)} = \ket{\Psi(t)}  -\beta(t) \ket{\Psi(0)}, \label{eq:CUSYGE}
\eeq
where $\beta(t)=\Braket{\Psi(0) | \Psi(t)}$.
Since the initial state is confined inside the potential barrier, this decaying state visualizes the out-going component.
\kimu{Then the decaying-density distributions,
\beq
\rho_{d}(t,\bir_1 ,\bir_2)= \abs{\Psi_d(t, \bir_1 , \bir_2)}^2, \label{eq:DAQUES}
\eeq
are presented in FIGs. \ref{fig:COKG_WS2_GS}-\ref{fig:COKG_WS2} (prompt case) and
in FIGs. \ref{fig:COKG_WS1_GS}-\ref{fig:COKG} (mixed case).
For plotting purpose, these functions are normalized at {\it each} point of time, namely, $\rho_d(t) \longrightarrow \rho_d(t)/ \Braket{\Psi_{d} (t) \mid \Psi_{d} (t)}$ for visualization.
}

\kimu{First we focus on the prompt case
with $R_b =14$ fm in FIG. \ref{fig:COKG_WS2_GS}.}
The decaying-density distribution is plotted for the variables similarly to Refs. \cite{2014Oishi, 2017Oishi, 2023Pfutzner_rev}.
Namely,
\beqa
&& r_{\rm p-p}= \sqrt{r^2_1 +r^2_2 -2r_1 r_2 \cos \theta_{12}},\nonumber \\
&& r_{\rm c-pp}= \frac{\sqrt{r^2_1 +r^2_2 +2r_1 r_2 \cos \theta_{12}}}{2}.
\eeqa
Here we remind that $r_{\rm p-p} \equiv \abs{\bir_2 -\bir_1}$ and $r_{\rm c-pp} \equiv \abs{\bir_2 +\bir_1}/2$.
\kimu{In FIG. \ref{fig:COKG_WS2}, the same plot but by changing the confining radius as $R_b =8.5$ fm is also displayed.
We mention this result in section \ref{sec:ddd}.
}

In FIG. \ref{fig:MAPS}, for the interpretation of \twop-decaying process, we illustrate the two typical processes, i.e., (i) the true \twop~emission \cite{2010Oishi, 2014Oishi} and (ii) the sequential $1p$-$1p$ emission.
Note that the mean opening-angle of two protons becomes narrow in the true-\twop~emission \cite{2010Oishi,2014Oishi}.

From FIGs. \ref{fig:COKG_WS2_GS} and \ref{fig:MAPS}, one can observe that
the prompt case exhibits the true \twop~emission as dominant process,
which is characterized by the asymmetric opening-angle distribution and the profile in the $r_1 \cong r_2$ region.
The dominance of true \twop~emission is naturally understood from the energy scheme.
Namely, in the prompt case, the $s_{1/2}$ resonance locates above the \twop~energy, and thus, the $1p$ emission is suppressed.
In FIG \ref{fig:COKG_WS1_GS} of the mixed case, in contrast, a sequential $1p$-$1p$ emission becomes apparent.
This appears as the trajectory along the $r_{c-pp} \cong r_{p-p}/2$ line.
This can be understood from the energy scheme:
since the intermediate $s_{1/2}$ resonance exists, the $1p$ emission can happen as the first step.

Comparing the prompt and mixed cases, the former one is consistent to experiments: the direct or true \twop~emission is experimentally confirmed for the ground-state resonance of $^{16}$Ne \cite{2014Brown, 2016Charity_EPJCON}.
Notice also that the ground-state width $\Gamma_1$ is smaller in the prompt case.
This is because the sequential emission via the $s_{1/2}$ resonance is more suppressed.
Thus, two protons need to escape mainly via the $d_{5/2}$ channel with the higher potential barrier.
Then the tunneling probability is reduced, and smaller width as well as longer lifetime are realized.

\subsection{second $0^+$ resonance} \label{sec:ddd}
\kimu{For finding the $0^+_{2}$ resonance,
as one numerical technique, we change the confining radius as $R_b = 8.5$ fm.
We confirmed that the initial-state properties do not remarkably change:
see FIGs. \ref{fig:GGG71_WS2} (prompt) and \ref{fig:GGG71_WS1} (mixed) for their density distributions, as well as
TABLE \ref{table:POX} for their energies, etc.
Then the time-dependent calculations are performed with the same $\hat{H}_{3B}$.
The calculated $P_{T}(t)$ are shown in FIGs. \ref{fig:2023_1113} and \ref{fig:2023_0618}.
}

\kimu{In FIGs. \ref{fig:2023_1113} and \ref{fig:2023_0618},
one can find a deviation from the pure-exponential damping.
Especially the rapid decrease of $P_{T}(t)$ followed by oscillation until $ct \cong 1200$ fm represents the mixture of $0^+_2$ resonance.
To interpret this deviation, we assume the superposition of the $0^+_{2}$ resonance with the $0^{+}_1$ one.
That is, for $\ket{\Psi(t=0)} =\int  F(E) \ket{E} dE$,
\beq
F(E) = \sqrt{N_{12}} \left[ L_1(E) + u_2 L_2(E) \right]^2,  \label{eq:TWOBW}
\eeq
where
\beqa
L_2 (E) &=& \sqrt{N_2} \frac{1}{E-(E_2 -i \Gamma_2 (E) /2)}, \nonumber  \\
\Gamma_2 (E) &=& \Gamma_2 \left[ 0.65 \left( \frac{E}{E_2}  \right)^2 +0.35 \left( \frac{E}{E_2}  \right)^4 \right].  \label{eq:Gofe2}
\eeqa
Namely, we assume the energy-dependent width as introduced in Ref. \cite{09Gri_80} for possibly the wide $0^+_2$ resonance.
Their normalization factors, $N_2$ and $N_{12}$, are determined so as to satisfy
$\int \abs{L_2 (E)}^2 dE = 1$ and
$\int S(E) dE = \int \abs{F(E)}^2 dE = 1$, respectively.
The parameter $u_2$ controls the relative amplitudes of the two $0^+$ resonances.
By fitting to the calculated $P_{T}(t)$,
we obtained the $0^+_2$-resonance parameters as summarized in TABLE \ref{table:VEY}.
These fitted functions, $P_{T}(t)$ and $S(E)$, are displayed in FIGs. \ref{fig:2023_1113} and \ref{fig:2023_0618}.
Notice that the energy distribution deviates from the single Breit-Wigner profile.
In TABLE \ref{table:VEY}, the $(s_{1/2})^2$ and $(d_{5/2})^2$ ratios of $0^+_2$ are also presented.
Those are evaluated by folding the initial state with the corresponding profile, $L_2 (E)$.
}

\kimu{From FIG. \ref{fig:2023_1113} and TABLE \ref{table:VEY} for the prompt case,
the $0^+_2$-amplitude parameter is fitted as $u_2 = -0.047$ when $R_b = 8.5$ fm.
The same results but with $R_b = 7.5$ fm are also presented, where $u_2 = -0.061$.
Namely, with the smaller $R_b$, the $0^+_2$ is more included.
We confirmed that, for $6 \leq R_b \leq 12$ fm, the common set of $0^+_2$-resonance parameters in TABLE \ref{table:VEY} are determined.
Thus, the $0^+_2$-resonance solution is suggested with the present $\hat{H}_{3B}$.
The same discussion applies to the mixed case.
}

\kimu{From TABLE \ref{table:VEY}, the $0^+_2$ resonance is $(s_{1/2})^2$ dominant.
From TABLE \ref{table:POX}, the inclusion of $0^+_2$ reduces the $(s_{1/2})^2$ ratio at $t=0$, whereas the $(d_{1/2})^2$ ratio is enhanced.
This is because the relative phase between $(d_{5/2})^2$ and $(s_{1/2})^2$ states in $0^+_2$ differs from that in $0^+_1$.}

The $0^+_2$ resonance is predicted at $Q_{2p} =2.96$ ($2.44$) MeV in the prompt (mixed) case.
The candidate of this state was reported in the pionic
double-charge-exchange reaction at $Q_{2p,~{\rm expt.}} =3.5$ MeV from \twop~threshold \cite{1997Fohl},
corresponding to the $E_x =2.1$ MeV excitation from the $0^+_1$ state.
This state is interpreted as the isobaric analogue to the $0^+_2$ state of $^{16}$C at $E_x=3.03$ MeV \cite{1977Fortune}.

\kimu{In FIGs. \ref{fig:COKG_WS2}
and FIGs \ref{fig:COKG},
the density-distributions of decaying states with $R_b = 8.5$ fm are presented.
By comparing the FIGs. \ref{fig:COKG_WS2_GS} and \ref{fig:COKG_WS2} in the prompt case,
an inclusion of the $0^+_2$ resonance helps the prompt-\twop~emission to become more apparent.
One explanation is given from the enhancement of $(d_{5/2})^2$ component, where the sequential emission is forbidden by the energy scheme.
In the mixed case in FIGs \ref{fig:COKG_WS1_GS} and \ref{fig:COKG}, in contrast,
the inclusion of $0^+_2$ however does not realize a dominant prompt-\twop~emission.
This is because the $(s_{1/2})^2$ component in this case allows the sequential emission.
}

\kimu{The connection between
the motion of particles and the energy spectrum is explained as follows.
Since the decaying state $\Psi_d(t,\bir_1 , \bir_2)$ has finite amplitudes
mostly in the outside region,
\beqa
\Braket{\Psi_d(t) \mid \Psi_d(t)} &=& \iint_{all} d\bir_1 d\bir_2 \rho_d(t,\bir_1,\bir_2)  \nonumber  \\
&\cong & \iint_{R_b \leq r_i} d\bir_1 d\bir_2 \rho_d(t,\bir_1,\bir_2).
\eeqa
On the other hand, $\Braket{\Psi_d(t) \mid \Psi_d(t)} = 1 - P_{T}(t)$ from its definition.
Thus,
\beqa
P_{T}(t) &=& \left| \int e^{-itE /\hbar} \abs{F(E)}^2 dE \right|^2  \nonumber  \\
&\cong & 1- \iint_{R_b \leq r_i} d\bir_1 d\bir_2 \rho_d(t,\bir_1,\bir_2).
\eeqa
This relation expresses that the superposition of two resonances, which is determined with the non-Breit-Wigner $\abs{F(E)}^2$, is connected with the decaying state.
}

%%%%%%%%%%%%%%%%%%%%%%%%%%%%%%%%%%%%%%%%%%%%%%%%%%%%%%%%%%%%%%%%%%%%%%%%%%%
\begin{table}[b] \begin{center}
\caption{
\kimu{Results for the $0^{+}_{1}$ and $0^{+}_{2}$~states of $^{16}$C.}
}
\label{table:PIVOT6}
\catcode`? = \active \def?{\phantom{0}} %define `?' as ' '(one-blank).
\begingroup \renewcommand{\arraystretch}{1.2}
\begin{tabular*}{\hsize} { @{\extracolsep{\fill}} lrrrr }
\hline \hline
&\multicolumn{2}{c}{prompt}    &\multicolumn{2}{c}{mixed}  \\
&$0^+_{1}$ &$0^+_{2}$ &$0^+_{1}$ &$0^+_{2}$  \\  \hline
$Q_{2n}=\Braket{\hat{H}_{\rm 3B}}$~[MeV] &$-5.484$ &$-0.860$         &$-5.430$ &$-3.022$ \\
$\Braket{v_{nn}}$~[MeV]     &$-6.199$  &$-0.584$                &$-4.765$ &$-1.081$ \\
$\Braket{v_{nn,Coul}}$~[MeV] &$0$    &$0$                &$0$ &$0$  \\
$(s_{1/2})^2$ ratio~[\%]  &$3.8$ &$96.0$                &$16.2$  &$82.9$  \\
$(d_{5/2})^2$ ratio~[\%]  &$92.9$ &$2.7$                &$79.9$  &$16.5$  \\
\hline \hline
\end{tabular*}
\endgroup
\catcode`? = 12 %initialize `?'.
\end{center} \end{table}
%%%%%%%%%%%%%%%%%%%%%%%%%%%%%%%%%%%%%%%%%%%%%%%%%%%%%%%%%%%%%%%%%%%%%%%%%%%

\subsection{Coulomb shift in $^{16}$Ne and $^{16}$C}
We give a notification on the $0^+_2$-$0^+_1$ gaps between
the mirror nuclei, $^{16}$Ne and $^{16}$C.
If the Coulomb force in $^{16}$Ne was neglected, their gaps should be equal.

\begin{figure}[t] \begin{center}
\includegraphics[width = 0.8\hsize]{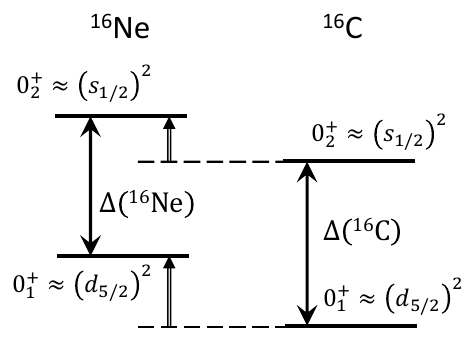}\\ (a) \\
\includegraphics[width = \hsize]{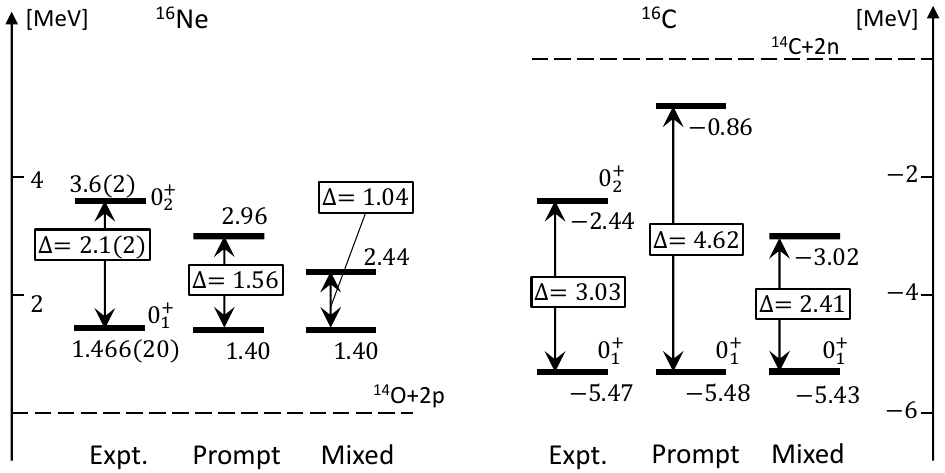}\\ (b) \\
\caption{(a) Schematic picture for the Coulomb shifts in the $^{16}$Ne and $^{16}$C, where
their $0^+_1$ ground states are assumed as $(d_{5/2})^2$-dominant.
\kimu{(b) Summary of $0^+_2$-$0^+_1$ gaps in the $^{16}$Ne and $^{16}$C nuclei in this work.
Experimental data are displayed for comparison \cite{1997Fohl, 1977Fortune}.}
}
\label{fig:WYFET_001}
\end{center} \end{figure}

First, in FIG. \ref{fig:WYFET_001}-(a),
we schematically assume that their $0^+_1$ ground states are $(d_{5/2})^2$-dominant.
Thus, their $0^+_2$ states are mostly of $(s_{1/2})^2$.
The coulomb force generally increase more the $d_{5/2}$ level, whereas
the $s_{1/2}$ level is less sensitive.
Thus, the $0^+_2$-$0^+_1$ gap of $^{16}$Ne should be smaller than that of $^{16}$C:
$\Delta({\rm ^{16}Ne}) < \Delta({\rm ^{16}C})$.
In FIG. \ref{fig:WYFET_001}-(b),
experimental data on these $0^+_2$-$0^+_1$ gaps
are summarized \cite{1997Fohl, 1977Fortune}.
One can find that $\Delta({\rm ^{16}Ne}) < \Delta({\rm ^{16}C})$.
Thus, the dominance of the $(d_{5/2})^2$ component in the $0^+_1$ state is expected.

%%%%%%%%%%%%%%%%%%%%%%%%%%%%%%%%%%%%%%%%%%%%%%%%%%%%%%%%%%%%%%%%%%%%%%
%%%%%%%%%%%%%%%%%%%%%%%%%%%%%%%%%%%%%%%%%%%%%%%%%%%%%%%%%%%%%%%%%%%%%%

\kimu{For checking the Coulomb shift,
we performed the same three-body calculations but for $^{16}$C ($^{14}$C$+n+n$).
Namely, we employ the same nuclear potentials for the valence two neutrons, switch off the Coulomb forces, and tune the additional-potential strength as
$w_0 = -442$ ($-141$) MeV$\cdot$fm$^3$ in the prompt (mixed) case for reproducing the ground-state energy.
Their results are summarized in TABLE \ref{table:PIVOT6}.
}

Our $0^+_1$ and $0^+_2$ energies for $^{16}$Ne and $^{16}$C are compared in FIG. \ref{fig:WYFET_001}-(b).
In the prompt case, $\Delta({\rm ^{16}Ne}) = 1.56$ MeV, which is smaller than the
corresponding gap of $^{16}$C, $\Delta({\rm ^{16}C}) =4.62$ MeV.
This is explained from the large (small) $d_{5/2}$ amplitudes
in the $0^+_1$ ($0^+_2$) states in the two nuclei.
Both the prompt and mixed cases are qualitatively consistent to the experimental situation.
In parallel,
our results especially of $^{16}$C quantitatively differ from experimental data.
Note that, in Ref. \cite{2015Gri_16Ne}, a breaking of isospin symmetry between the two systems is suggested.
In such a case, further optimizations of nuclear potentials are necessary.
This work is left for forthcoming studies.

For the $s_{1/2}$ and $d_{5/2}$ amplitudes, a sizable difference exists between our results and other works \cite{2015Gri_16Ne, 06Horiuchi}.
In Ref. \cite{2015Gri_16Ne}, the \twop-emission of $^{16}$Ne is $s_{1/2}$-dominant.
There, the $d_{5/2}$ and $s_{1/2}$ amplitudes are shown as essential to reproduce the $0^+$ and $2^+$ energies.
The same dominance of $s_{1/2}$ is concluded in Refs. \cite{2015Gri_16Ne, 06Horiuchi} for the $0^+_1$ state of $^{16}$C.

\section{summary} \label{sec:summary}
In this work, the \twop-radioactive emission from the meta-stable $0^+$ state of $^{16}$Ne is investigated with the time-dependent three-body model.
The \twop-decaying width of the $0^+_1$ resonance is of keV-order and beyond the present reach of measurement, due to the instrumental nature.
By observing the time-dependent decaying state in the prompt case, a dominance of true emission is confirmed.
This is a natural product of the energy scheme with the high $s_{1/2}$ resonance \cite{2003Peters}.
On the other hand, in the mixed case by assuming the intermediate $s_{1/2}$ resonance, the sequential emission appears as expected.

\kimu{The $0^+_2$ resonance is solved by using the technique of modified confining potential, where the initial state can have a finite inclusion of the $0^+_2$ resonance.
In the prompt case, the $0^+_2$ resonance is predicted at $\cong 3$ MeV from the \twop~threshold.
In addition, by including this $0^+_2$ resonance, the true \twop~emission becomes more dominant.
This is one product of the interference of two resonances.
}

Search for certain emission process, where the predicted $0^+_2$ resonance can be apparent, is in progress now.
One possible case was reported in the pionic double-charge-exchange reaction \cite{1997Fohl}, where its energy is $\cong 1$ MeV higher than our result.
On the other side,
in the one-neutron-knockout reaction from $^{17}$Ne to experimentally produce $^{16}$Ne, the $0^+_2$ resonance is not expected to be involved \cite{2014Brown}.

This work is limited to the $0^+$ configuration.
In Refs. \cite{2002Grigorenko, 2015Gri_16Ne}, the Thomas-Ehrman shift (TES) between the $0^+$ and $2^+$ resonances of $^{16}$Ne and its isobaric-analogue states in $^{16}$C has been investigated.
This TES can be utilized for the model optimization involving the uncertain $s_{1/2}$ resonance in $^{15}$F.
The TES is not regarded in this work, due to that the numerical costs of time-dependent calculations for the $2^+$ states should be expensive.
Instead, we consider the ambiguity of $s_{1/2}$ resonance in $^{15}$F by comparing the two cases.

Checking the sensitivity to the initial state, on which the experimental production history of $^{16}$Ne should be reflected, is waiting for further investigation \cite{2021Wang_Naza, 2023SMWang}.
For comparison with experimental data on \twop~correlations, the final-state interactions are fully taken into account.
Several computational problems remain for this purpose.

\begin{acknowledgments}
T. Oishi especially thanks Chong Qi, Simin Wang, and Futoshi Minato for fruitful discussions.
Numerical calculations are supported by the cooperative project of supercomputer Yukawa-21 in Yukawa Institute for Theoretical Physics, Kyoto University.
We appreciate the Multi-disciplinary Cooperative Research Program (MCRP) 2023-2024 by Center for Computational Sciences, University of Tsukuba (project ID wo23i034), allocating computational resources of supercomputer Wisteria/BDEC-01 (Odyssey) in Information Technology Center, University of Tokyo.
\end{acknowledgments}


%---
%%\bibliographystyle{apsrev4-1}
%%\bibliography{zb_all}

%%%\newpage
%%%(To be continued.)
%%%\input{POWDAM_Append.tex}

\end{document}